\documentclass[conference]{IEEEtran}

\usepackage[ruled,vlined]{algorithm2e}
\usepackage{algorithmic}        % set algorithm font size
\DeclareUnicodeCharacter{2212}{-}

\usepackage{amsmath}    % math equation

\usepackage{graphicx} % allowed to modify the images
\usepackage{fancyhdr} % create sub-image and sub-caption

\usepackage[labelfont=bf, skip=5pt, font=small]{caption}
\usepackage{subcaption} 

\usepackage{array} % table vertical central
\usepackage{makecell} % table make cell position

\usepackage{cite}

\author{\IEEEauthorblockN{Junyao Zhang}
\IEEEauthorblockA{\textit{University of Southern California} \\
Los Angeles, CA \\
junyaozh@usc.edu}
\and
\IEEEauthorblockN{Paul Bogdan}
\IEEEauthorblockA{\textit{University of Southern California} \\
Los Angeles, CA \\
pbogdan@usc.edu}
\and
\IEEEauthorblockN{Shahin Nazarian}
\IEEEauthorblockA{\textit{University of Southern California} \\
Los Angeles, CA \\
shahin.nazarian@usc.edu}
}

\begin{document}
\title{A Majority Logic Synthesis Framework For Single Flux Quantum Circuits}
\maketitle

\begin{abstract}
Exascale computing and its associated applications have required increasing degrees of efficiency.
Semiconductor-Transistor-based Circuits (STbCs) have struggled with increasing the GHz frequency while dealing with power dissipation issues. 
Emerging as an alternative to STbC, single flux quantum (SFQ) logic in the superconducting electrons (SCE) technology promises higher-speed clock frequencies at ultra-low power consumption.
However, its quantized pulse-based operation and high environmental requirements, process variations and other SFQ-specific non-idealities are the significant causes of logic error for SFQ circuits.
A suitable method of minimizing the impact of the afore-mentioned error sources is to minimize the number of Josephson Junctions (JJs) in the circuits, hence an essential part of the design flow of large SFQ circuits.
This paper presents a novel SFQ logic synthesis framework that given a netlist, offers an automated mapping solution including majority (MAJ) logic with the goal of minimizing the number of JJs, while catering to the unique characteristics and requirements of the design. Our experiments confirm that our synthesis framework significantly outperforms the state-of-the-art academic SFQ technology mapper, namely reducing the number of JJs on average by 35.0\%.
\end{abstract}

\section{Introduction}
As a solid competitor to the state-of-the-art semiconductor circuits in the field of super electronic products, SCE has shown great potential in ultra-high-speed clock frequency, and ultra-low power consumption \cite{likharev1991rsfq}.
The interest in SFQ has grown in communities that need higher performance and higher energy efficiency as the most touted candidate in SCE. 
However, the high sensitivity of SFQ circuits to the environmental factors has been problematic \cite{likharev1991rsfq, bunyk2001rsfq}. There has been some research on hard faults in post-manufacturing testing \cite{gaj1995parameter,vernik1999experimental}.
However, the SFQ technologies have much larger feature sizes compared to CMOS technologies \cite{narasimha20177nm}. E.g., a recent fabrication process of JJs is based on a 200nm production-level \cite{tolpygo2014fabrication}. A large feature size results in low defect density, thereby stuck-at and other hard faults become less critical.
% However, SFQ technology is not physically scaling-focused. It achieves high operating speed and low power consumption by using unique characteristics of Josephson Junctions and superconducting operations. Although the fabrication process of JJs based on 200nm production-level has been developed \cite{tolpygo2014fabrication}, it is still a much larger feature size compared to CMOS technology \cite{narasimha20177nm}. Larger feature size means low defect density, thereby stuck-at and other hard faults become less critical.

In contrast, other non-idealities such as process variations and other SFQ-specific issues (inductive coupling, bias current steering, flux trapping) have significant impacts on the operation of the fabricated chips \cite{wang2019automatic}.
These faults can cause the logic gates to produce erroneous outputs just as highly-distorted pulses can not be interpreted logically correctly by cells, or unexpected jitters can produce a logical HIGH \cite{celik2012statistical}. A variation-focused ATPG paradigm is proposed to generate test patterns for these random faults \cite{wang2019automatic}.
However, these faults still can not be eliminated or minimized effectively. Therefore, it is critical to develop an effective approach that can mitigate the negative impact of these sporadic failures. 

Logic synthesis is an essential step in the design flow of digital circuits because it enables the desired metrics to be minimized \cite{hachtel2007logic}. 
It is divided into two distinct phases: \textbf{1) technology-independent optimizations:} Boolean optimizations such as restructuring, re-substitution, common sub-expression extraction, and node minimization. 
\textbf{2) technology mapping:} associating logic expressions with actual gates in the given cell library.
According to the theoretical and experimental results \cite{bunyk2001rsfq,  wang2019automatic}, the aforementioned faults can be effectively reduced by minimizing the number of Josephson Junctions (referred to as the \#JJs) in the SFQ circuits. Thus, an effective method of mitigating the effects of non-idealities in SFQ circuits is to use a mapping algorithm that is oriented toward reducing \#JJs. 

Additionally, logic synthesis can be improved by incorporating majority logic (MAJ) gates into expressions rather than relying exclusively on the AND-OR-INV-based (AOI) representation \cite{amaru2014majority}.
Although MAJ circuit configuration is proposed in Dynamic Single Flux Quantum (DSFQ) \cite{krylov2020asynchronous}, and MAJ logic synthesis works exist in both CMOS \cite{amaru2015majority} and Adiabatic Quantum-Flux-Parametron (AQFP) \cite{cai2019majority}. MAJ circuit configuration in SFQ and a systematic synthesis framework capable of integrating MAJ in to the expression is still imminent.

 \begin{figure*}[h]
    \centering
    \setlength{\abovecaptionskip}{0.1cm}
    \vspace{-0.4cm}
    \includegraphics[width=0.8\textwidth]{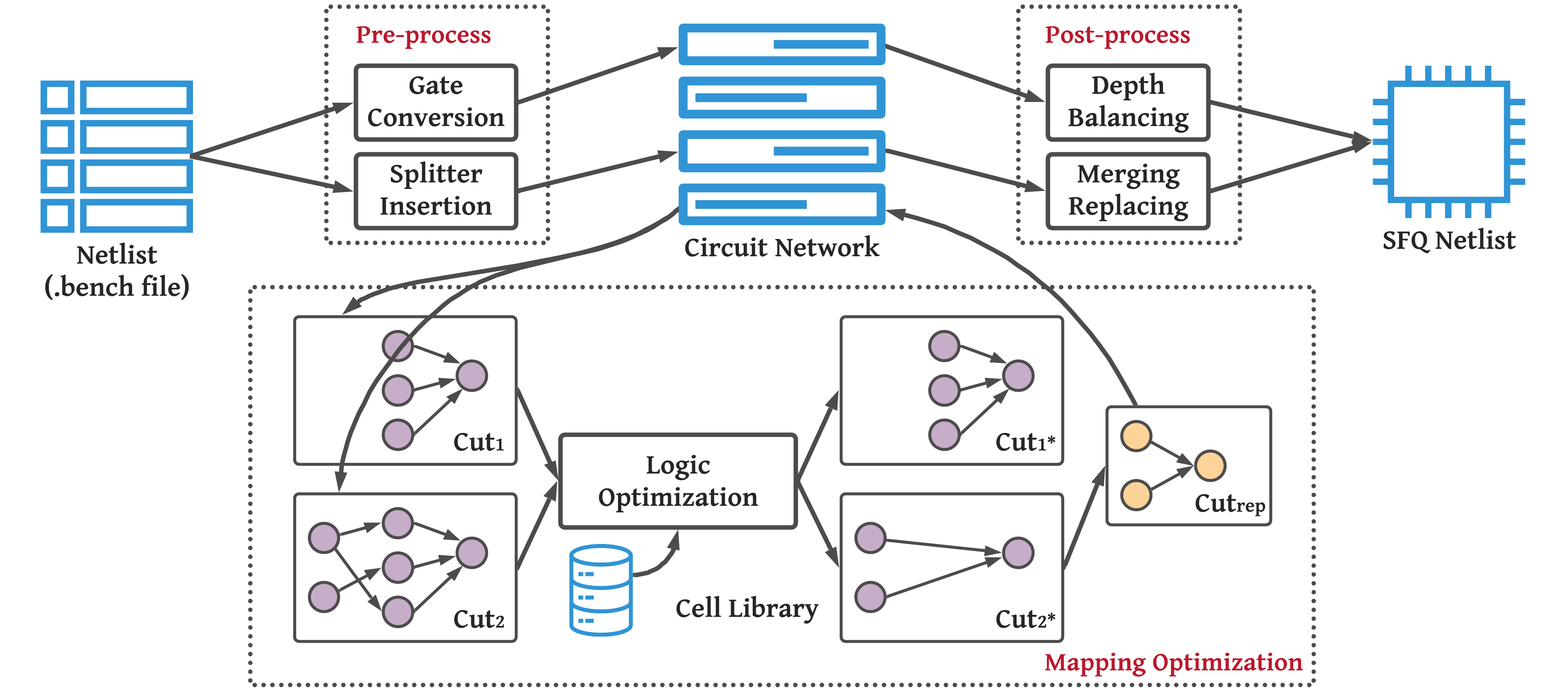}
    \caption{\textbf{Overview of the proposed framework}.
\textbf{1) Pre-process:} insert splitters into the input netlist; convert all of the cells in the netlist to the cells in the cell library.
\textbf{2) Mapping optimization:} 
find all the $K$-feasible cuts for each node;
optimize the Boolean function of each cut circuit $Cut_n$ with the modified Quine-McCluskey algorithm;
regenerate the cut circuit $Cut_n*$ with its optimized Boolean function;
modify this node by representative cut $Cut_{rep}$ which is the regenerated cut circuit with the most significant improvement in PND value).
\textbf{3) Post-process:} insert DFFs to balance the logic level of each node's fan-in; merge and replace the path-depth balancing DFFs to further minimize their overhead in the SFQ netlist.} \label{fig:flow}
\end{figure*}

In this paper, we propose a logic synthesis framework for generating MAJ included SFQ netlists from a given netlist.
Mapped netlist minimizes the number of JJs and thus the risk of encountering non-idealities in SFQ circuits.
The framework, as illustrated in Fig. \ref{fig:flow}, is divided into three distinct phases:
\textbf{1) pre and post-process:} both processes are related to the circuit modifications unique to SFQ, such as path-depth balancing, splitter binary trees insertion, and gate conversion. \textbf{2) mapping optimization:} optimize the cut circuits for each node with a modified Quine-McCluskey algorithm, then greedily modify each node with its corresponding representative cut to reduce the product of total \#JJs and the network depth of the netlist (denoted as PND). 
The main contributions can be summarised as follows:
\begin{itemize}
    \item We propose a circuit configuration of MAJ gate in SFQ logic and apply it to our proposed logic synthesis framework.
    \item We develop a novel greedy mapping algorithm for SFQ logic, which can modify each node with its representative cut to reduce PND value.
    \item We propose a logic synthesis framework for SFQ logic, which can process the netlist to an SFQ netlist, including catering to the unique characteristics of SFQ logic and mitigating the impact of non-idealities in SFQ circuits
\end{itemize}
The rest of the paper provides the preliminaries, related work, motivation, cell characterization under non-idealities, majority gates, synthesis framework for SFQ logic, evaluation, and conclusion in successive sections.

\section{Preliminaries}
\subsection{SFQ Logic}
Overdamped Josephson Junctions in SFQ use resistive current bias, which means that the binary information is present in a picosecond duration of voltage pulses $V(t)$, rather than voltage levels in semiconductor technologies. Due to the voltage pulse is equivalent to a quantum of flux $\phi_o= 2.07{\times}10^{−15} V{\cdot}s$, it is also referred to as single flux quantum pulse (SFQ pulses) These unique properties fundamentally alter the conventional understanding of the representation of bits. The basic convention is that the arrival of an SFQ pulse during the current clock period represents a logic “$1$” whereas the absence of any pulse during this period is understood as logic “$0$”. Additionally, there are several SFQ circuit-specific properties that need to be discussed below.

\subsubsection{Fan-out}\label{sec:sp} 
\begin{figure}[b]
    \centering
    \setlength{\abovecaptionskip}{0.1cm}
    \vspace{-0.4cm}
        \includegraphics[width=0.45\textwidth]{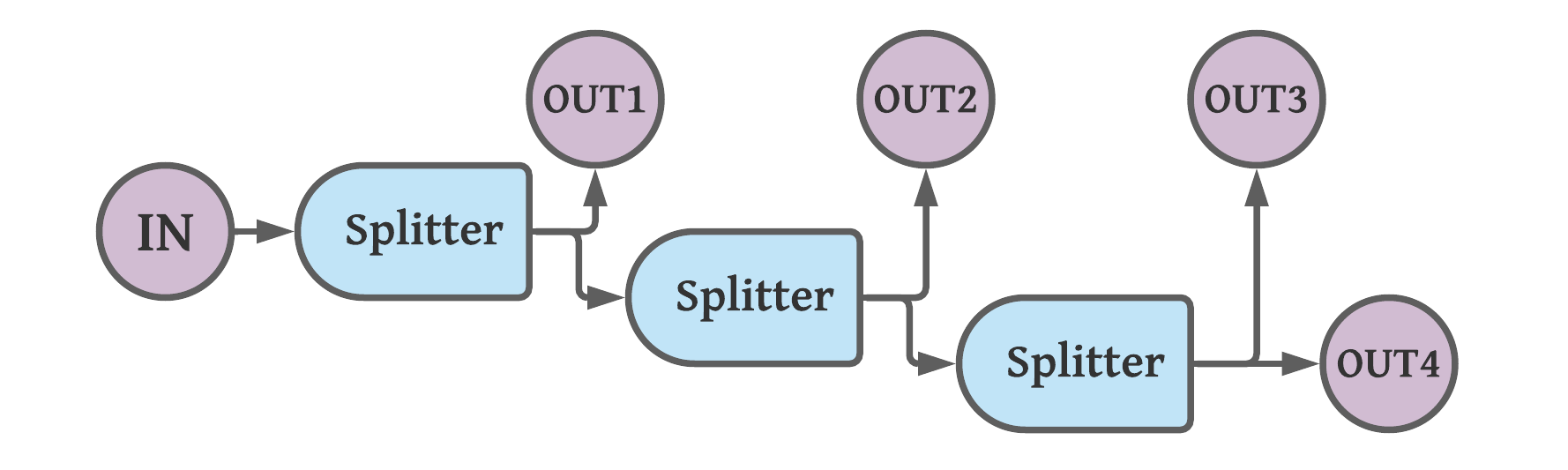}
        \caption{A splitter binary tree (Fan-outs = 4)}
        \label{fig:SPT}
\end{figure}

Given the quantum communication nature of SFQ logic, the fan-out of cells strictly equals to one. A special asynchronous SFQ gate (splitter cell) is used to achieve higher fan-out \cite{likharev1991rsfq}. A splitter can generate only two output pulses after accepting an SFQ pulse. Additional splitters can be applied to a binary tree structure to allow for additional fan-outs. It needs $n-1$ splitters to accommodate $n$ fan-outs. Fig. \ref{fig:SPT} is an example of the splitter binary tree when fan-outs equal 4.

\subsubsection{Gate-level Pipeline}\label{sec:pip} Almost all SFQ gates (except for some asynchronous gates like splitter or Josephson Transmission Line) are clock-synchronized, so an SFQ gate can be thought of as a CMOS gate with an edge-triggered flip-flop at its output. The outputs only respond to their respective inputs when the clock pulse arrives. This work \cite{friedman2001clock} outlines the design of a clock distribution network, and its take-away is that the SFQ circuit must be completely gate-level pipelined.

\subsubsection{Path-depth Balancing}\label{sec:pdb} Due to the gate-level pipeline prerequisite, all inputs of an SFQ gate must have the same logic level for correct operation. Some D flip-flops (DFFs) should be inserted into the path to balance the logic level of inputs if there is a difference among logic levels of fan-ins. Assume that the fan-in depths of a OR gate are different: $in_1 = 1$, $in_2 = 2$. The correct pulse as $in_1$ will be consumed 1 clock earlier than $in_2$ without the path-depth balancing DFF. Thus, path-depth balancing DFFs are needed to ensure the correct logic values.

\subsection{Logic Synthesis}\label{sec:sys}
There are two phases in logic synthesis: the technology-independent and the technology-dependent phase \cite{hachtel2007logic}. Several optimizations are performed in the first stage to reduce the total number of literals in the given network. Some functional operations include common subexpression extraction, decomposition, and resubstitution \cite{rudell1989logic}.
The second phase is referred to as technology mapping, in which appropriate gates from a given library are allocated to nodes in the given network in order to satisfy certain constraints, such as minimizing certain cost functions \cite{hachtel2007logic}.
The provided netlist needs to be transformed into a Boolean network prior to technology mapping. Boolean networks are directed acyclic graphs (DAG) \cite{hachtel2007logic}, in which nodes represent logic gates and directed edges correspond to the wires that link the gates. The terms network and circuit are used interchangeably in this paper.
Primary inputs (PIs)/primary outputs (POs) are nodes without fan-ins/fan-outs in the current network \cite{mishchenko2007combinational}. 
A cut of node $n$ is specified as follows: $n$ is the root node (the output of this cut), $l_n$ is a set of nodes in the network referred to as leaf nodes (inputs of this cut) \cite{mishchenko2007combinational}. Each path from a PI to $n$ passes through at least one leaf. A cut is $K$-feasible if it contains no more than $K$ leaf nodes \cite{cong1994flowmap}. 
The logic level of a node is the length of the longest path from any PI to the node. The logical depth is the largest level of an internal node in the network.

\section{Related Work}
In the literature of the logic synthesis for SFQ circuits, a few papers have presented the technology-independent and technology mapping problems \cite{yamashita2006transduction, pasandi2018pbmap, pasandi2018sfqmap, katam2017design}. 
In \cite{yamashita2006transduction}, a framework is developed by constructing a virtual cell named “2-AND/XOR”, which enables the use of the CMOS logic synthesis tools for SFQ circuits.
In \cite{katam2017design}, it added path-depth balancing DFFs and the splitters to the netlist followed by applying the standard retiming algorithm \cite{leiserson1991retiming} to reduce the required number of path-depth balancing DFFs. 
In \cite{pasandi2018sfqmap}, a technology mapping tool for SFQ logic circuits (called SFQmap) is presented, with an emphasis on minimizing logical depth and product of the worst-case stage delay and the logical depth (PSD). 
\cite{pasandi2018pbmap} is the similar work to SFQmap, which can generate mapping solutions with balanced structures. 
Mapping algorithms in \cite{pasandi2018sfqmap, pasandi2018pbmap} are implemented by applying Cut-enumeration-based technology, which is close to the work in \cite{cong1994flowmap}.
The above SCE domain papers  \cite{pasandi2018pbmap, pasandi2018sfqmap, katam2017design} are all developed on an AND-INV graph (AIG) based logic synthesis and verification tool, ABC \cite{mishchenko2007abc}.

Furthermore, there are several applications beyond CMOS technologies like nanotechnologies that have benefited from MAJ logic synthesis, such as quantum cellular automata (QCA) \cite{lent1997device} and single electron tunnelling (SET) \cite{oya2002majority}. In SCE domain, there are few papers that mention majority logic. One of them proposed the majority logic synthesis framework for AQFP \cite{cai2019majority}, another proposed an asynchronous Dynamic SFQ Majority Gates design \cite{krylov2020asynchronous}.

\section{motivation}\label{sec:motivation}
Despite significant research advances that leverage logic synthesis in a different domain, the state-of-the-art synthesis tool in SFQ logic still has a substantial improvement space. In \cite{pasandi2018pbmap, pasandi2018sfqmap}, their mapping algorithm is developed on Cut-enumeration-based technology. 
This technology aims to map the cut circuit with the supergate by look-up table (LUT). 
Specifically, the mapping algorithm establishes a massive supergate library and searches it for each cut circuit. If a supergate has the same LUT as the cut circuit, then this supergate can be used to map the cut circuit.
The drawbacks of this technology are the limited $K$ value and the memory overhead associated with the supergate library, which grows exponentially with the logical depth and basic cell library size. For instance, by having 20 gates in the original library and using up to level 3 supergates, there will be around 4000 supergates. 
Despite a large-scale supergate library is established, it is still hard to find a mapped supergate capable of implementing the function of a cut as $K$ value increases \cite{pasandi2018pbmap}. 
Also, the aforementioned technology primarily focused on the optimization of the depth and path-depth balancing overhead. Apart from these optimizations, we wish to integrate majority logic into the mapping algorithm in order to decrease the \#JJs and thus mitigate the effects of random faults.

We propose a novel SFQ logic synthesis framework that favors generating mapping solutions with lower impact SFQ based non-idealities by reducing JJ numbers. Several closed-form formulae are developed for: cut selection, modified Quine-McCluskey algorithm, and netlist processing algorithms with SFQ-specific properties. To the best of our knowledge, this is the first paper that presents a logic synthesis framework in SFQ logic: 1) involving majority gates, and 2) mitigating the effect of non-idealities in SFQ circuits.

\section{Cell Characterization Under Non-idealities}\label{sec:cell}
SFQ circuits achieve fast operating speeds and low power consumption by using unique characteristics of JJs and superconducting operations.
However, JJs are highly sensitive to environmental conditions. Some SFQ-specific issues like inductive coupling, bias current steering, flux trapping can significantly impair their operation \cite{likharev1991rsfq, bunyk2001rsfq}. 
In \cite{wang2019automatic}, Monte Carlo simulations are used to characterize the behaviors of SFQ logic cells in non-idealities, and the average value is normalized using Gaussian distributions when circuit parameters change. 
The simulation results demonstrate that the error rates are relatively higher when a cell contains more Josephson Junctions. 
As a consequence, SFQ circuits with a large number of JJs are more likely to encounter errors due to non-idealities. 
As illustrated in Section \ref{sec:pip}, the latency of an SFQ circuit is primarily determined by its logical depth. While a netlist can mitigate the effects of non-idealities by minimizing the \#JJs, it does not expect to achieve this objective at the expense of other critical circuit properties, such as logical depth. Therefore, we introduce a new metric, PND, to evaluate latency and reliability combinational performance of a mapping solution. For a network $\mathcal{C}$ with node set $N$, PND is defined as the product of the total \#JJs and its logical depth $d_\mathcal{C}$.
\begin{align}\label{eqn:PND}
P(\mathcal{C}) = (\sum_{n\in N} J_n) \times d_\mathcal{C}
\end{align}
where P() is the PND value, and $J_n$ is the \#JJs for node $n$. The cell library used in this paper consists of the following cells, and each cell is listed in this format: \textit{cell name (\#JJs)}.

Cell library: DFF ($8$), AND2 ($9$), AND3 ($12$), AND4 ($15$), OR2 ($9$), OR3 ($11$), OR4 ($13$), INV/inverter ($5$), XOR ($7$), SP/splitter ($3$), and the proposed MAJ ($12$).

% Specifically, the occurrence of any above non-idealities issue for a Josephson Junction represents that the cell of this Josephson Junction will perform an error output.
% On the contrary, if the number of Josephson Junctions in an SFQ circuit is reduced effectively, which enhances its reliability significantly.

 \begin{figure}[h]
    \centering
        \setlength{\abovecaptionskip}{0.1cm}
    \vspace{-0.4cm}
        \includegraphics[width=0.4\textwidth]{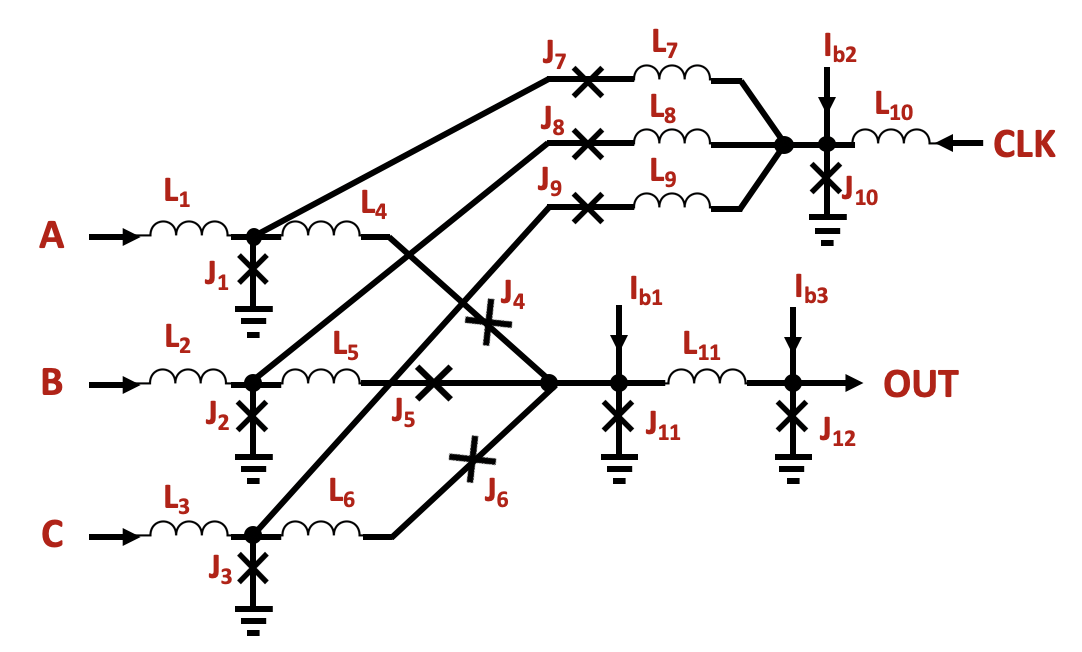}
        \caption{Circuit configuration for majority gate}
        \label{fig:MAJ}
\end{figure}

\section{Majority Gate Model}\label{sec:MAJ}
The majority gates are discussed in this section, along with an example circuit configuration of a gate suitable for use in large-scale SFQ circuits. The n-input (n being odd) majority function M returns the logic value assumed by more than half of the inputs \cite{sasao2012switching}. 
The principle of Josephson Junction is that when the value of its bias current approaches its critical current, it can induce a $2\pi$ leap at the voltage phase, and then trigger an SFQ pulse. 
A proposed three-input SFQ majority gate is shown in Fig. \ref{fig:MAJ} The gate consists of three parallel superconducting quantum interferometer loops (SQUID) \cite{likharev1991rsfq}, $S1: \{J1, L4, J4, J11\}$, $S2:\{J2, L5, J5, J11\}$ and $S3:\{J3, L6, J6, J11\}$. These loops terminate on a grounded junction J11 that drives the output port OUT. 

\begin{figure}[h]
    \centering
    \setlength{\abovecaptionskip}{0.1cm}
    \vspace{-0.4cm}
        \includegraphics[width=0.45\textwidth]{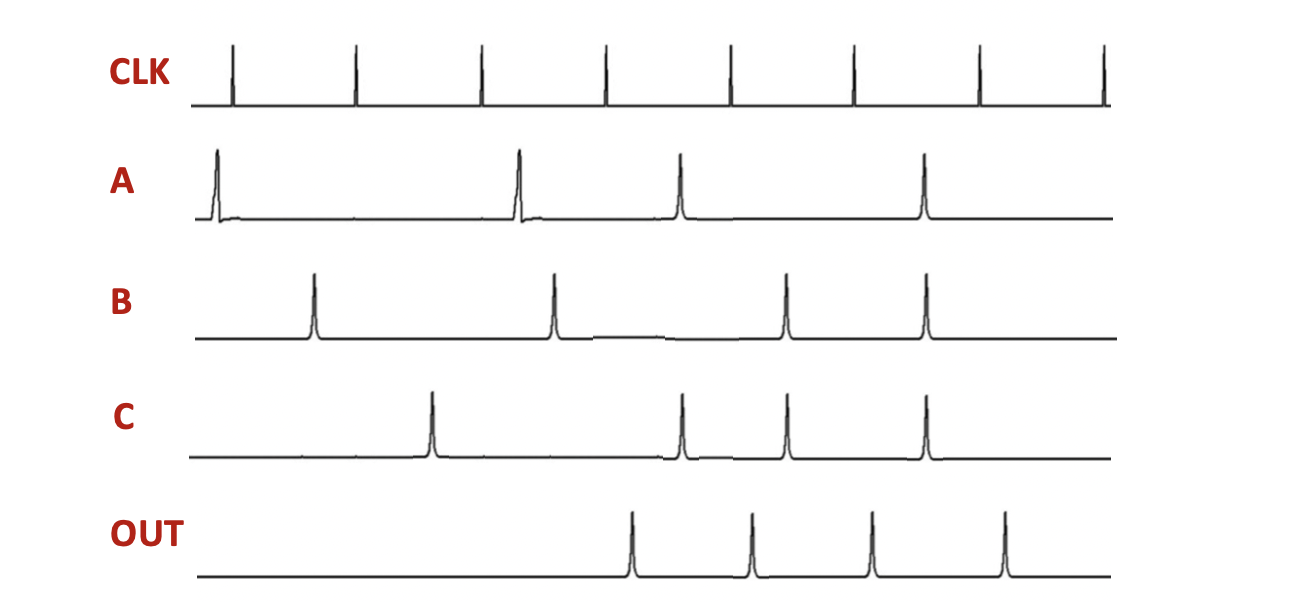}
        \caption{Operation of the proposed 3-inputs majority gate, $A$, $B$ and $C$ are the inputs, $CLK$ is the clock input, and $OUT$ is the output}
        \label{fig:sim}
\end{figure}

The proposed 3-input MAJ gate operates in a similar manner to the SFQ 3-input AND gate \cite{katam2018logic}. However, compared with AND3, the dc-bias $I_{b1}$ in majority gate provides a comparatively large current. When input pulses are fed in a single clock cycle, the associated loops leap to the '1' state, and the total bias current at J11 equals $I_ {b1}$ plus currents in leaped loops.
In the initial state, all loops are initialized to '0'. 
In case, there is only one input pulse or no pulse in a single clock cycle, the bias current in J11 does not reach its critical current.
Therefore, the clock pulse is incapable of inducing a $2\pi$ leap at J11. When the cell is supplied with two or more input pulses, the total bias current approaches or exceeds to its critical value as the dc-bias $I_{b1}$ provides a larger current. In this situation, the clock pulse will induce an SFQ pulse at J11, resulting in a '1' at $OUT$. The simulation of the proposed gate is depicted in Fig. \ref{fig:sim}, and its operation is represented as 
\begin{equation}
M(A,B,C) = AB + BC + AC \notag
\end{equation}
% If there is only one input pulse or no pulse is fed in a single clock cycle. The bias current in J11 does not reach its critical current. 

 \begin{figure}[t]
    \centering
    \setlength{\abovecaptionskip}{0.1cm}
    \vspace{-0.4cm}
        \includegraphics[width=0.45\textwidth]{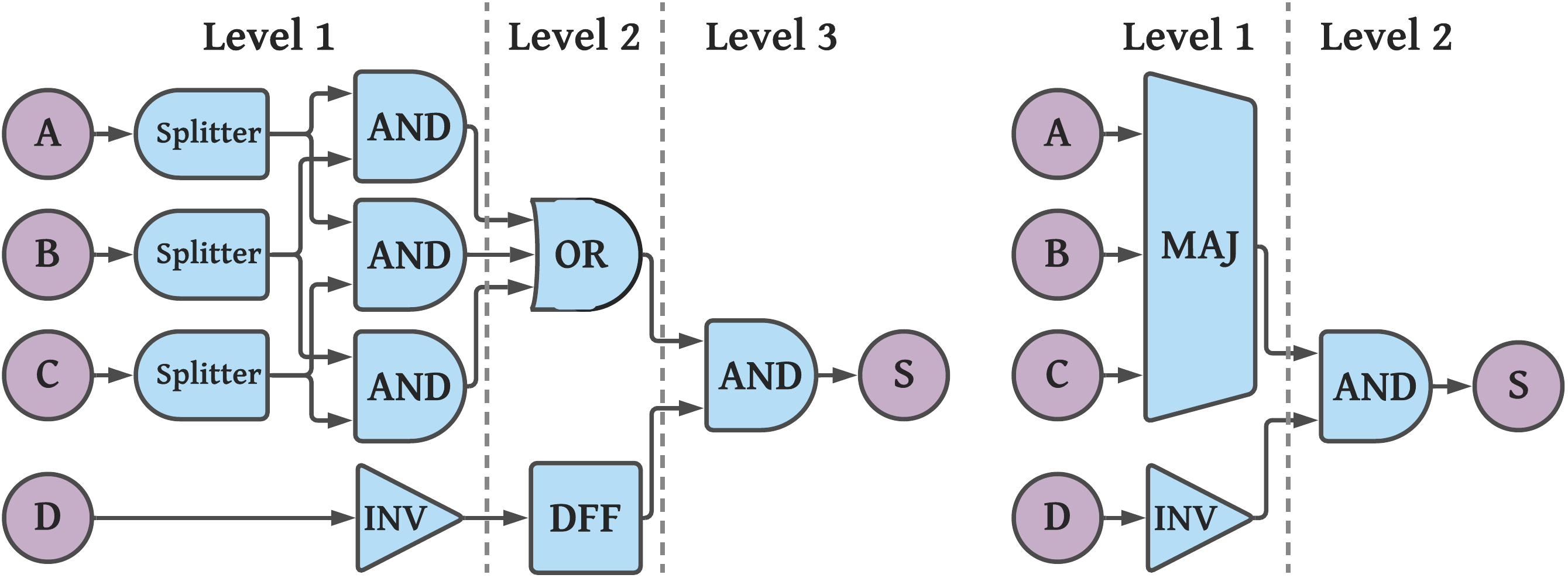}
        \caption{Two mapping solutions for $S=(ab + bc + ac)\overline d$. The one on left is generated by SFQmap \cite{pasandi2018sfqmap}, which PND value is 201, maximum logic level is 3, and requires 1 path-depth balancing DFF; the one on right can reduce the above 3 metrics to 52, 2, 0 separately.}
        \label{fig:example}
\end{figure}

\section{Synthesis Framework For SFQ Logic}\label{sec:framework}
For a given Boolean expression: $S=(ab + bc + ac)\overline d$. We hope to map it into an SFQ circuit. As illustrated in Fig. \ref{fig:example}, cutting-edge mapper (SFQmap \cite{pasandi2018sfqmap}) generate the left circuit, which has a PND value of 201 and a logical depth of 3. 
However, it is possible to have a better mapping solution with a lower PND value and improved logic optimization, as shown in the right graph in Fig. \ref{fig:example}. 
The reason for this is that no algorithm is currently implemented in state-of-the-art technology mappers for minimizing \#JJs and encompassing MAJ gates, let alone optimize such a multi-objective value as PND.
Our framework can generate novel mapping solutions with minimal PND values and path-depth balancing overhead. 
Suppose for a network $\mathcal{C}$ with node set $N$, $\mathcal{K}_n$ is the set of all $K$-feasible cuts of node $n\in N$, and $c^{j*}_n$ is the regenerated cut circuit of cut $c^j_n \in \mathcal{K}_n$, then we formalize the objective for network $\mathcal{C}$ with node set $N$ as:
\begin{equation}\label{eqn:obj}
\min (P(\sum_{\forall n\in N} (c^{rep}_n)))
\end{equation}
% &\min (\sum_{n\in N} P(c^{j*}_n) \quad | \quad d^{j}_n = \max \limits_{\forall I^i_n \in \mathcal{K}_n}(d^{i}_n)  
% \\ 
% &\min (\sum_{n\in N} pb(n))
% where $pb(n)$ is the number of required Josephson Junctions for node $n$ in the path-depth balancing process,
where $P()$ is the PND value function in Eq. (\ref{eqn:PND}). In the following section, we will first elaborate on the greedy mapping algorithm for the proposed PND value minimization problem, followed by theoretical analysis for the mapping algorithm, and processes of SFQ unique characteristics modifications in our framework.

\subsection{Mapping Optimization}\label{sec:map}
In our framework, we perform the following heuristic greedy approach to reduce the PND value of the network. The optimal result for mapping a network is defined as the result that iterates through all the nodes in the network and modifies each node with its representative cut circuit. The representative cut $c^{rep}_n$ is the one that has the greatest decrease in PND value after regeneration from the optimal Boolean function compared with original cut $c^{j}_n$. The approach is broken down into the following sub-tasks.

\begin{figure}[h]
    \centering
    \setlength{\abovecaptionskip}{0.1cm}
    \vspace{-0.4cm}
        \includegraphics[width=0.45\textwidth]{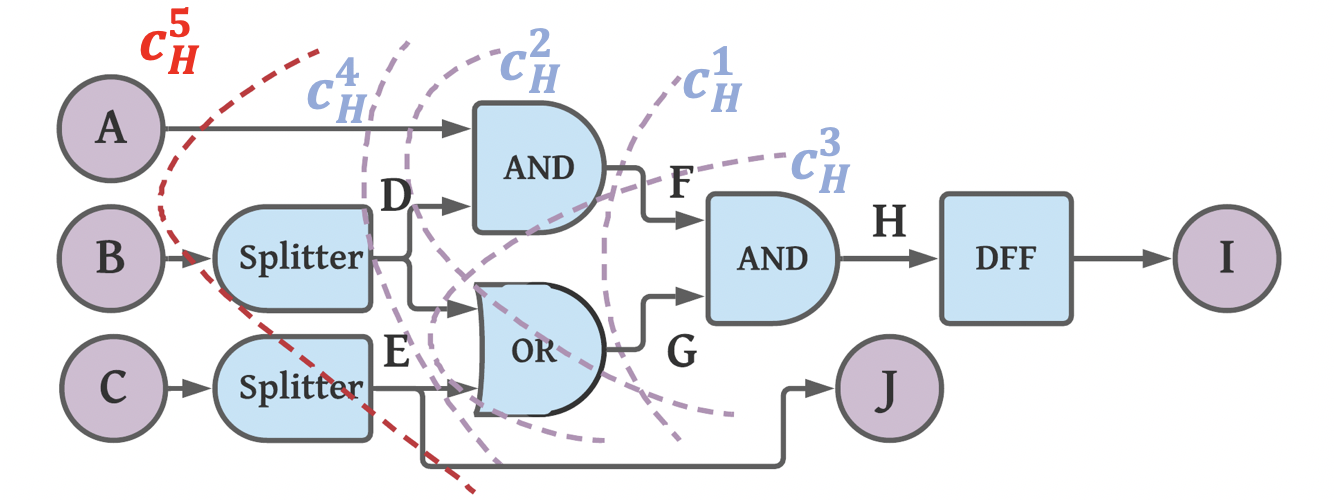}
        \caption{3-feasible cuts for node H, $c^1_H$: \{F, G\}, $c^2_H$: \{A, D, G\}, 
        $c^3_h$: \{D, E, F\}, $c^4_h$: \{A, D, E\}, $c^5_H$: \{A, B, E\}}
        \label{fig:cut}
\end{figure}

\textbf{(1) Find $\mathcal{K}_n$:} We explain $K$-feasible cut in Section \ref{sec:sys}. Algorithm \ref{alg:kcut} performs a Breath-First search for the upstream nodes of root nodes $n$ and collects all the possible cut circuits under the $K$ boundary, specifically, leaf nodes number of each cut $c^j_n$ is in range $[2, K]$. Suppose $I^j_n$ denotes the set of leaf nodes for cut $c^j_n \in \mathcal{K}_n$. 
In general, the searching terminates at the PI nodes, splitter nodes or when the number of leaf nodes exceeds $K$ (size$(I^j_n)>K$), since the regenerated cut circuit $c^{j*}_n$ retains only the same leaf nodes and root node as its preceding cut circuit $c^{j}_n$. If $c^{j}_n$ contains a splitter node inside, and some branches of this splitter are omitted. The latter regeneration process has an effect on the logic values in non-included branches, as this node can be removed or reconnected. Our algorithm is heuristic that can detect possible cuts even behind splitters. Cut $c^5_H$ in Fig. \ref{fig:cut} is an example.

\begin{algorithm}
% \algsetup{linenosize=\tiny}
  \scriptsize
\caption{Find all $K$-feasible cuts}\label{alg:kcut}
\SetAlgoLined
\SetKwInOut{Input}{Input}
\SetKwInput{KwOutput}{Output}  
\Input{root node $n$, cut parameter $K$}
\KwOutput{set $\mathcal{S}$ of all leaf nodes sets for each cut $c_j$}
    $\mathcal{U}$: fan-in set; $N$: node set; g(): gate type\\
    set curPI = $\{n\}$, idx = 0 \\
    \While{$size(curPI) > idx$}{
    $n_c$ = curPI[idx]\\
    s = size(curPI) + size($\mathcal{U}_{n_c}$) - 1\\
    add all $\mathcal{U}_{n_x}$ to $N_{in}$, $(n_x \in curPI)$\\
    \For{each $ i \in N_{in}$}{
        \uIf{$\exists j = i; j\in N_{in}, \mathcal{U}_a; i \in \mathcal{U}_b; a, b\in$curPI}{
            add $\mathcal{U}_a, \mathcal{U}_b-i$ to $N_{x}$ and add $a$, $b$ to $N_{d}$\\
            remove i, j from $N_{in}$
        }
    }
    \uIf{curPI-$N_{d}$ + $N_{x}\leq$ k}{
        curPI = curPI - $N_{d}$ + $N_{x}$\\
        add curPI to $\mathcal{S}$\\
    }\uElseIf{g$(n_c)\notin$ \{PI/Splitter\} and s$<K$}{
        add $\mathcal{U}_{n_c}$ to curPI, remove $n_c$ from curPI\\
        add curPI to $\mathcal{S}$\\
    }\uElse{
        idx++\\
    }
    % \If{$n_x \in \mathcal{U}_{n_a}, \mathcal{U}_{n_b}$ and $n_a, n_b \in$ curPI}{
    %     \If{idx + size$(\mathcal{U}_{n_a})$+ size$(\mathcal{U}_{n_b})-1\leq K$ }{
    %         add $\mathcal{U}_{n_a}$, $\mathcal{U}_{n_b}$ to curPI ($n_x$ only add once);\\
    %         add curPI to $\mathcal{S}$\\
    %         continue;\\
    %     }
    % }
    }
\end{algorithm}

\textbf{(2) Find optimal Boolean function for each cut $c^j_n \in \mathcal{K}_n$:} In Section \ref{sec:motivation}, we discuss the defects of the state-of-the-art mapping algorithm. We present a novel Boolean function minimization algorithm based on the Quine-McCluskey algorithm \cite{mccluskey1956minimization} to overcome their defects.
Our optimization algorithm is superior since it incorporates XOR and MAJ logic into the Boolean Algebra.
In addition, in a number of majority-logic-based works \cite{cai2019majority, amaru2014majority, amaru2015majority}, all the AND/OR gates are substituted by with MAJ gates. 
Although these majority logic based mapping algorithms have demonstrated significant performance gain, and Section \ref{sec:MAJ} proves our proposed MAJ gate has the same overhead as the 3-input AND gate. They can not imply that the MAJ is the panacea for all logic optimizations. 
% Although these majority-inverter graph (MIG) based algorithms have demonstrated significant performance gains over AOI \cite{amaru2014majority}, and Section \ref{sec:MAJ} proves our proposed MAJ gate has the same overhead as the 3-input AND gate. They can not imply that the MAJ is the panacea for all logic optimizations. 
For example, 3 MAJs and 2 INVs are needed to represent an XOR function. If we insert the splitters and do the path-depth balancing, the overhead will further surge. We only involve the cuts with MAJ gates in the certain best situations in our framework.

To begin the logic optimization, we simulate the cut $c^j_n$ to determine all of its prime implicants. Assume that the cut $c^j_n$ has three leaf nodes: \{a, b, c\} and each bit is represented by $0$ or $1$. We present the logic with the standard representation, sum of product (SOP) \cite{de1994synthesis}. Assume the implicant set is as follow:
\begin{equation}
  c^j_n = \{111, 110, 101, 011, 001, 000\}\notag
\end{equation}
Implicant \{$110$\} represents $c^j_n=1$, when a = 1, b = 1, and c = 0. The number and order of bits in each implicant are immutable.
Then, our algorithm evaluates the relationship between each prime implicant and combines them to generate the essential prime implicants. The final set of the essential prime implicants is the optimal Boolean function of cut in the form of SOP. The following relationships in Algorithm \ref{alg:Boolean} are used to validate the implicants: grey code pair, XOR/XNOR logic and MAJ logic.
Grey code pair check examines whether there is exactly one different bit between two implicants, and replaces this complement bit with don't care '-' sign. 
MAJ logic check compares three implicants simultaneously to replace the bits in these implicants that can perform majority logic with '$\star$'. XOR/XNOR logic check is similar to MAJ logic check with two implicants each time. '$\oplus$' denotes the bits that have an XOR relationship, while '$\ominus$' denotes XNOR.
The combination begins with grey code pair check until all the implicants are essential prime implicants in this checking stage, then the output implicants are then subjected to other two logic checks (MAJ and XOR/XNOR). 
These two checks are mutually exclusive, in that if an implicant is modified by one check while the other ignores this implicant.
Our algorithm reduces the preceding implicant set to the following essential prime implicants. 
\begin{equation}
  c^{j}_n = \{11-, 1-1, -11, 00-\} = \{\star\star\star, 00-\}\notag
\end{equation}
These implicants represent the optimal Boolean function as:
\begin{equation}
  c^{j}_n = M(a,b,c) + \overline{ab} \notag
\end{equation}

\begin{algorithm}[h]
\algsetup{linenosize=\tiny}
\scriptsize
\caption{Implicant checks}\label{alg:Boolean}
\SetKwInput{KwInput}{Input}                % Set the Input
\DontPrintSemicolon
\KwInput{implicant string s1, s2, s3}
% fan-in set $\mathcal{F}_{n_i}$ and  fan-out set $\mathcal{D}_{n_i}$ of node $n_i$
% Set Function Names
\SetKwFunction{FDB}{Greycode check}
\SetKwFunction{FSBTI}{XOR/XNOR logic check}
\SetKwFunction{FGC}{MAJ logic check}

    \SetKwProg{Fn}{Function}{:}{\KwRet}
    \Fn{\FDB{$s1, s2$}}{
        $flag = 0$, $idx = 0$ \\
        \For{each bit $i\in s1$}{
            \If{$s1[i] \neq s2[i] $}{
                $flag$++, $idx = i$       \\
            }
        }
        \If{$flag == 1$}{
          \KwRet merge s2 to s1 and s1[$idx$]$\rightarrow -$\\
        }
    }
 
    \SetKwProg{Fn}{Function}{:}{}
    \Fn{\FSBTI{$s1, s2$}}{
        f1 = 0, f2 = 0, $s$ = \{\}\\
        \For{each bit $i\in s1$}{
            \If{$(s1[i],s2[i]) == (0,1)$}{
                $f1$++, add i into $s$ \\
            }
            \If{$(s1[i],s2[i]) == (1,0)$}{
                $f2$++, add i into $s$ \\
            }
        }
        \If{$f2 == 2$ or $f2 == 2$}{
            \KwRet merge s2 to s1 and for $j\in s$ s1[j]$\rightarrow \ominus$\\
            }
        \If{$f1==1$ and $f2 == 1$ }{
          \KwRet merge s2 to s1 and for $j\in s$ s1[j]$\rightarrow \oplus$\\
        }
        
    }
    
    \SetKwProg{Fn}{Function}{:}{}
    \Fn{\FGC{$s1, s2, s3$}}{
        $f1 = 0$, $f2 = 0$, $f3 = 0$, $s$ = \{\}\\
        \For{each bit $i\in s1$}{
            \If{$(s1[i],s2[i],s3[i]) == (1,1,-)$}{
                $f1$++, add i into $s$ \\
            }
            \If{$(s1[i],s2[i],s3[i]) == (1,-,1)$}{
                $f2$++, add i into $s$ \\
            }
            \If{$(s1[i],s2[i],s3[i]) == (-,1,1)$}{
                $f3$++, add i into $s$ \\
            }
        }
        \If{$f1 == 1$ and $f2 == 1$ and $f3 == 1$}{
            \KwRet merge s2,s3 to s1 and for $j\in s$ s1[j]$\rightarrow \star$\\
        }
    }
\end{algorithm}

\textbf{(3) Regenerate $c^{j*}_n$ from the optimal Boolean function of $c^{j}_n$:}
$K$ value only limits the number of leaf nodes in the $K$-feasible cut searching, which means that each leaf node can exist at a variety of logic levels. 
Considering $c^2_H$ in Fig. \ref{fig:cut}, the regenerated cut circuit $c^{2*}_H$ is a 3-input AND gate from $H = ADF$, the optimal Boolean function of $c^{2}_H$. It seems to make a dramatic improvement in PND value by lowering it to 12. 
Nevertheless, $c^{2*}_H$ increases the overhead in Josephson Junctions as 2 DFFs are required to balance the path-depth of nodes A and D. 
The PND value surges to 56 which is even higher than $c^{2}_H$ (42). Therefore, it is essential to regenerate $c^{j*}_n$ by taking into account the logic levels of leaf nodes in $c^{j}_n$.

\textbf{(4) Compare the PND value of $c^{j*}_n$ with $c^{j}_n$, update $c^{rep}_n$, and modify the node $n$ with its representative cut $c^{rep}_n$:} 
The regenerate cut $c^{j*}_n$ and cut $c^{j}_n$ signify the a cut pair. The improvement in each pair is calculated using the following equation.
\begin{equation}\label{eqn:imp}
    d^{j}_n = P(c^{j}_n) - P(c^{j*}_n)
\end{equation}
where $d^{j}_n$ is the improvement value of PND value between $c^{j}_n$ and $c^{j*}_n$, and P() is the PND values computed by Eq. (\ref{eqn:PND}).
If $d^{j}_n$ in has the largest value among all cut pairs ($c^{i}_n$, $c^{i*}_n$), $c^{i}_n \in \mathcal{K}_n$, which indicates $c^{j*}_n$ has the greatest positive improvement in PND value compared with $c^{j}_n$, and then $c^{j*}_n$ is updated to the representative cut $c^{rep}_n$. 
If none of the $c^{j*}_n$ can improve the PND value for node $n$, the $c^{rep}_n$ is set to null. 
The node $n$ is modified with its representative cut $c^{rep}_n$ to minimize the PND value of network $\mathcal{C}$ after all its cuts $c^{j}_n \in \mathcal{K}_n$ has been traversed. 
If none of the nodes in the network are modified, this indicates that the network is already the optimal mapping solution after the pre-process stage.
% This demonstrates that our algorithm is greedy due to the worst-case scenario is that none of the nodes in the network are modified, implying that the network is already the optimal mapping solution after the pre-process stage. 

% 1) Compute the representative cut of node $n$:
% \begin{align}
% I^{rep}_n = I^{i+}_n, \max \limits_{\forall I^i_n \in \mathcal{K}_n}(P(I^i_n)-P(I^{i+}_n)) 
% \end{align}
% &\min (\sum_{n\in N} P(c^{j*}_n) \quad | \quad d^{j}_n = \max \limits_{\forall I^i_n \in \mathcal{K}_n}(d^{i}_n) 
% 2) Update each node with its representative cuts:
% \begin{align}
% \mathcal{M}^+ = f(I^{rep}_n) \quad \forall n \in N
% \mathcal{K}_n}(P(I^i_n)-P(I^{i+}_n)))
% \end{align}
% \begin{align}
% \mathcal{M}^+ = f(I^{rep}_n) \quad \forall n \in N
% \end{align}

\subsection{Theoretical Analysis}
The optimal solution for mapping the network $\mathcal{C}$ is formalized with the following equations:\\
1) Compute the representative cut $c^{rep}_n$ of node $n$:
\begin{equation}\label{eqn:rep}
    c^{rep}_n = c^{j*}_n\quad | \quad d^{j}_n = \max \limits_{\forall c^i_n \in \mathcal{K}_n}(d^{i}_n)
\end{equation}
2) Update each node with its representative cut:
\begin{equation}\label{eqn:map}
    \mathcal{C}= \sum_{\forall n\in N} (c^{rep}_n) 
\end{equation}
\textbf{Theorem 1.} The objective function Eq. (\ref{eqn:obj}) is submodular.\\
\\\textit{proof.} For a given network $\mathcal{C}$ with node set $N$, we define two sets of nodes can generate representative cuts by our framework, $\alpha\subseteq \{a_1, \dots, a_n\}$, $\beta\subseteq \{b_1, \dots, b_n\}$ where both sets $\alpha, \beta \subseteq \Omega$ and $\Omega$ is the solution space of the problem, and we define $f(x)$ as the objective function in Eq. (\ref{eqn:obj}).
If $\alpha\cap \beta = \epsilon$, there exists $\epsilon = \emptyset$.
\begin{align}
&f(\alpha\cap \beta) + f(\alpha\cup \beta)        \notag\\
&= \sum_{n\in N}\{ |P(\alpha(n))| + |P(\beta(n))| - |P(\epsilon(n))| + |P(\epsilon(n))|\}  \notag\\
&= \sum_{n\in N}|P(\alpha(n))| + \sum_{n\in N}|P(\beta(n))|   \notag\\
&= f(\alpha) + f(\beta) \notag
\end{align}
On the basis of the preceding equation, we can deduce that the objective function is submodular. Due to for any two sets $\alpha, \beta \subseteq \Omega$, $f(\alpha)+f(\beta) = f(\alpha\cap \beta) + f(\alpha\cup \beta)$.\\

\noindent\textbf{Theorem 2.} The objective function Eq. (\ref{eqn:obj}) is monotonic.\\
\\\textit{proof.} For a network $\mathcal{C}$ with node set $N$, we define an arbitrary node set can generate representative cuts by our framework $\alpha'=\alpha\cup \tau$, we can get $f(\alpha\cup \tau) - f(\alpha) \geq 0$.
\begin{align}
f(\alpha\cup \tau) 
&= \sum_{n\in N}\{ |P(\alpha'(n))|           \notag\\
&= \sum_{n\in N}\{ |P(\alpha(n))| + |P(\tau(n))| \}  \notag\\
&= \sum_{n\in N}|P(\alpha(n))| + \sum_{n\in N}|P(\tau(n))|   \notag\\
&= f(\alpha) + f(\tau) \notag
\end{align}

\subsection{Pre-Process}
We introduce our mapping algorithm in the previous section. However, the framework needs to pre-process the netlist before the mapping optimization by interpreting the netlist to an SFQ circuit-based network whose nodes represent the logic gates in the netlist. These two steps are needed in the pre-process stage of this representation scheme: \textit{1) Gate conversion:} convert all the gates in the netlist to the existing cells in the cell library. Assume the input netlist contains NAND gates that are not included in our cell library. Therefore, each NAND gate is converted to an AND gate and a new INV node is concatenated. In a similar fashion, the conversion function modifies XNOR and NOR gates. In addition, due to the cell library's constraint, the AND/OR gate's fan-ins can not exceed four. The higher fan-in gates are transformed into gate trees by their lower fan-in counterparts; \textit{2) Splitter insertion:} Section \ref{sec:sp} presents the fan-out characteristic of the SFQ circuits. A splitter binary tree must be inserted into each node with a fan-out greater than two. Algorithm \ref{alg:pre-process} includes extensive descriptions of the gate conversion and splitter insertion.
\begin{algorithm}[h]
\algsetup{linenosize=\tiny}
\scriptsize
\caption{SFQ circuit unique characteristics modifications}
\label{alg:pre-process}
\SetKwInput{KwInput}{Input}                % Set the Input
\DontPrintSemicolon
\KwInput{Nodes set $N$ for Network $\mathcal{C}$, }
% fan-in set $\mathcal{F}_{n_i}$ and  fan-out set $\mathcal{D}_{n_i}$ of node $n_i$
% Set Function Names
\SetKwFunction{FDB}{Path-depth Balancing}
\SetKwFunction{FSBTI}{Splitter Insertion}
\SetKwFunction{FGC}{Gate Conversion}
$\mathcal{U}$: fan-ins; $\mathcal{D}$: fan-outs; $l$: logic level; g(): gate type\\
    \SetKwProg{Fn}{Function}{:}{\KwRet}
    \Fn{\FDB{$N$}}{
        \For{each node $n \in N$}{
            \If{g$(n)\neq \textbf{splitter}$ }{
                $l_{max}$ = $\max(l_j, j\in\mathcal{U}_n)$\\
                \For{each fan-in $i \in\mathcal{U}_n$}{     
                    insert ($l_{max}-l_i$) DFFs at $i$
                }
            }{}
        }
    }
 
    \SetKwProg{Fn}{Function}{:}{}
    \Fn{\FSBTI{$\mathcal{N}$}}{
        \For{each $n \in N$}{
            \If{g$(n)\neq\textbf{splitter}$ and $\mathcal{D}_n>1$ }{
                $c = n$\\
                \For{each fan-out $o_j \in\mathcal {D}_n$}{
                    % $j\in(1,...,size({D}_{n_i})-1)$
                    insert a \textbf{splitter} $s^j$ between $c$, $o_j$\\ 
                    create edge $c\rightarrow s^j$, $s^j\rightarrow o_j$\\
                    \If{$j=size({D}_n)-1$}{
                    create edge $s^j\rightarrow o_{j+1}$
                    }
                    $c = s^j$
                }
            }{}
        }
    }
    
    \SetKwProg{Fn}{Function}{:}{}
    \Fn{\FGC{$\mathcal{N}$}}{
        \If{g$(n)\in \{$NAND/NOR/XNOR$\}$ }{
            insert $n_*$ between $n$ and its fan-out $\mathcal{D}_n$\\
            g($n_*$) $\in$ \{AND/OR/XOR\}, g($n$) $\in$ INV
        }
        \If{g$(n)\in \{$AND/OR$\}$ and size $(\mathcal{U}_n)>4$}{
            insert $n_*$ between $n$ and $\mathcal{D}_n$, g($n_*$)$\in$\{AND/OR\}\\
            % create edge $ n_{**} \rightarrow n_{i*}$ (g$(n_{**})\in$\{AND/OR\})\\
            \For{$f_j\in\mathcal{U}_n$ and $j\leq4$}{
                move edge $f_j\rightarrow n_*$ 
            }
        }
    }
  
\end{algorithm}

\subsection{Post-Process}
In the post-process, the framework needs to modify the network with property in Section \ref{sec:pdb}: all fan-ins of each gate should have the same delay (clock phases). This modification is delegated to the process behind the optimization in order to prevent duplicate cut searching and computation. As a result of path-depth balancing, DFFs have the same cuts as their fan-in nodes by considering the nodes I and H in Fig. \ref{fig:cut}. The pseudo-code of path-depth balancing is also listed in Algorithm \ref{alg:pre-process}.

 \begin{figure}[h]
    \centering
    \setlength{\abovecaptionskip}{0.1cm}
    \vspace{-0.4cm}
        \includegraphics[width=0.45\textwidth]{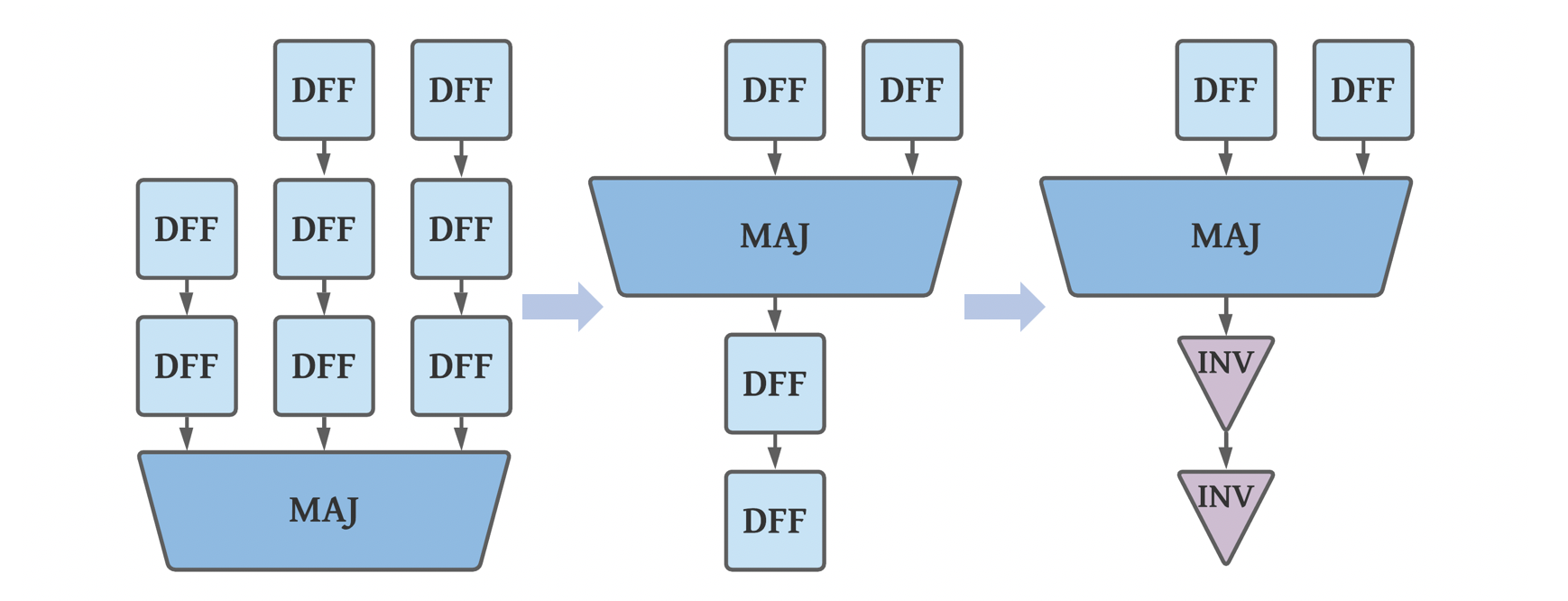}
        \caption{Merging \& replacing example, fan-ins of target gate has 2, 3, 3 DFFs, respectively. This process can reduce \#(DFFs+INVs) amount by 50\% in this specific case.}
        \label{fig:merge&replace}
\end{figure}
Additionally, since DFF consumes more \#JJs than INV, a merging \& replacing process is used to further minimize path-depth balancing DFFs consumption.
The framework only processes the DFFs inserted by the path-depth balancing algorithm in this stage, the DFFs that are contained in the input netlist are not changed to retain the correct logic function of the network.
For each cell, if all of its fan-ins have at least $y$ DFFs inserted, $y$ is the shared number of DFFs for this gate. So all its fan-ins can remove $y$ DFFs, and then insert $y$ DFFs to its fan-outs without logic change.
In the meantime, for each path, if there are $x$ path-depth balancing DFFs inserted ($x > 2$), the $x-(x\%2)$ DFFs is replaced with INVs.
As shown in Fig. \ref{fig:merge&replace}, three fan-ins to the target gate have 2, 3, and 3 DFFs, respectively. This structure can be simplified by relocating two DFFs to the target gate's fan-out path, reducing the number of DFFs (\#DFFs) at the fan-ins to 0, 1, and 1, respectively, and then replacing the DFFs at the fan-out path with INVs. This step ensures that the total number of path-depth balancing DFF is as low as possible while also mitigating the effects of non-idealities in the circuit.

\begin{table}[b]
    \centering
    \caption{Experimental Results for proposed framework in different $K$ values. $3^-$ denotes $K=3$ and the merging \& replacing process is disabled.}
    \label{tab:diff}

    \begin{tabular}
    {m{1.2cm}<{\centering} m{1.2cm}<{\centering} m{0.9cm}<{\centering} m{1.5cm}<{\centering} m{1.8cm}<{\centering} }
    \hline\hline
    $K$           & Logical Depth   & \#JJs     & PND($\times10^5$) &\#(DFFs+INVs)  \\    [0.8ex]\hline 
    $K=3^-$     &46.7           &108662     &63.48              &11247          \\    [0.8ex]\hline
    $K=4^-$     &44.8           &106253     &62.42              &10943          \\    [0.8ex]\hline 
    $K=3$       &46.1           &81401      &45.71              &11032          \\    [0.8ex]\hline
    $K=4$       &44.8           &80038      &45.0               &10733          \\    [0.8ex]\hline
    \end{tabular}
\end{table}

\begin{algorithm}
\footnotesize
\caption{Mapping optimization}\label{alg:mapping}
\SetAlgoLined
\SetKwInOut{Input}{Input}
\SetKwInput{KwOutput}{Output}  
\Input{Network $\mathcal{C}$, cut parameter $K$}
\KwOutput{Mapped circuit $\mathcal{C}_{map}$}
$l$: logic level; $I$: leaf nodes set; P(): PND value\\
    \For{each node $n\in N$}{
        compute all $K$-feasible cuts $\mathcal{K}_n$ for $n$ with $K$\\
        $d$ = 0, Representative cut $c^{rep}_n$ = NULL;\\
        \For{each cut $c^j_n\in \mathcal{K}_n$}{
            compute optimal Boolean function $F_j$ of cut $c^j_n$\\
            generate $c^{j*}_n$ from $F_j$ with $l_{I^j_n}$\\
            \If{$P(c^j_n)-P(c^{j*}_n) > d$}{
            $d = P(c^j_n)-P(c^{j*}_n)$, $c_{rep} = c^{j*}_n$
            }
        }
        \If{$c^{rep}_n\neq$ NULL}{
            modified the $\mathcal{C}$ by cut $c^{rep}_n$ 
        }
    }
\end{algorithm}

\begin{table*}[t]
    \centering
    \caption{Experimental Results for proposed framework $\mathcal{F}$ ($K=4$) and baselines ($\mathcal{B}_1:$ Input netlists, $\mathcal{B}_2:$ SFQmap \cite{pasandi2018sfqmap})}
    \label{tab:result}

    \begin{tabular}
    {m{1.1cm}<{\centering} m{0.9cm}<{\centering} m{0.9cm}<{\centering} m{0.9cm}<{\centering} 
     m{0.9cm}<{\centering} m{0.9cm}<{\centering} m{0.9cm}<{\centering}
     m{0.9cm}<{\centering} m{0.9cm}<{\centering} m{0.9cm}<{\centering}
     m{0.9cm}<{\centering} m{0.9cm}<{\centering} m{0.9cm}<{\centering} }
    \hline\hline
    {} & \multicolumn {3}{c}{Logical Depth} & \multicolumn {3}{c}{\#JJs}
    & \multicolumn {3}{c}{PND($\times10^5$)} & \multicolumn {3}{c}{\#(DFFs+INVs)}\\
   Circuits & $\mathcal{B}_1$ & $\mathcal{B}_2$ & $\mathcal{F}$ 
   & $\mathcal{B}_1$ & $\mathcal{B}_2$ & $\mathcal{F}$ 
   & $\mathcal{B}_1$ & $\mathcal{B}_2:$ & $\mathcal{F}$ 
   & $\mathcal{B}_1$ & $\mathcal{B}_2:$ & $\mathcal{F}$\\   \hline 
    b04     &51  &40 &33   &35297   &30141  &18532  &18.0  &12.1  &6.12  &3855   &3294  &2332 \\    [0.8ex]\hline
    b11     &61  &35 &36   &48190   &31251  &22636  &29.4  &10.9  &8.15  &5480   &3267  &3137 \\    [0.8ex]\hline 
    c880    &42  &35 &23   &19764   &14551  &8380   &8.3   &5.1   &1.93  &2172   &1566  &1096 \\    [0.8ex]\hline
    c1355   &44  &14 &12   &33995   &10915  &7587   &15.0  &1.53  &0.91  &3792   &1024  &896  \\    [0.8ex]\hline
    c1908   &68  &52 &50   &70512   &58294  &25983  &47.9  &30.3  &13.0  &8436   &6773  &3963 \\    [0.8ex]\hline
    c2670   &42  &38 &33   &30768   &29475  &21998  &12.9  &11.2  &7.3   &3205   &2970  &2601 \\    [0.8ex]\hline
    c3540   &75  &64 &58   &65143   &58443  &39179  &48.9  &37.4  &22.7  &7215   &6357  &5320 \\    [0.8ex]\hline
    c5315   &73  &53 &52   &114273  &91323  &66002  &83.4  &48.4  &34.3  &12783  &9784  &9257 \\    [0.8ex]\hline
    c6288   &246 &168 &168 &402824  &248038 &166283 &991   &416.7 &279.4 &47935  &28874 &28159\\    [0.8ex]\hline
    c7552   &68  &49 &48   &127514  &106264 &79554  &86.7  &52.7  &38.2  &13998  &10981 &10209\\    [0.8ex]\hline
    s1423   &67  &59 &57   &82400   &55746  &30214  &55.2  &32.9  &17.2  &9723   &6518  &5174 \\    [0.8ex]\hline
    s1494   &18  &16 &15   &19328   &20447  &14721  &3.48  &3.27  &2.21  &1722   &1774  &1569 \\    [0.8ex]\hline
    s5378   &33  &30 &26   &58191   &55137  &37101  &19.2  &16.5  &9.65  &6490   &6097  &4354 \\    [0.8ex]\hline
    s35932  &42  &16 &13   &1121734 &310597 &239681 &471.1 &49.7  &31.2  &126866 &26715 &24724\\    [0.8ex]\hline
    s38584  &70  &63 &48   &815009  &727571 &422713 &570.5 &458.3 &202.9 &89538  &79792 &58209\\    [0.8ex]\hline
    Avg  &66.7&48.8&44.8   &202996  &12312  &80037  &164.1 &79.1  &45.0  &22881  &13017 &10773\\    [0.8ex]\hline
    % s1238   &32  &27 &29   &26741   &24570  &17921  &8.5   &6.6   &5.19  &2844 &2560 &2495 \\  [0.8ex]\hline
    % c499    &12  &12 &12  &5835    &5835   &4395   &0.7   &0.7   &0.52  &544  &544 &544 \\      [0.8ex]\hline 
    % s9234   &73  &53 &x   &192702  &159755 &32  &140.6 &84.7  &256  &18479 &15184 &129 \\\hline
    % s13207  &71  &56 &x   &171963  &158496 &32  &122.1 &88.7  &256  &13360 &13191 &129 \\\hline
    \hline
    Imp(Avg)&$\uparrow$32.8\% &$\uparrow$8.2\%  &{} &$\uparrow$60.5\% &$\uparrow$35.0\% &{}  
            &$\uparrow$72.5\% &$\uparrow$43.1\% &{} &$\uparrow$53.1\% &$\uparrow$17.5\% &{}\\    [0.8ex]\hline
    \end{tabular}
\end{table*}

\section{Evaluation}
In this section, we provide two experiment setups and experimental results to validate the effectiveness of our proposed framework, several ISCAS arithmetic benchmark circuits \cite{brglez1989combinational} for testing our developed framework 
The pseudo-code of our proposed framework is shown in Algorithm.\ref{alg:mapping}.
First, we present different $K$ value experiments, where we compare the performance of different $K$ values, as well as corresponding results without merging \& replacing processes. Then we compare our proposed framework and the state-of-the-art technology mapper.

\textbf{1) $K$ values test:} In Section \ref{sec:map}, we present the $K$-feasible cut finding algorithm and mapping optimization algorithm. We are now conducting an empirical evaluation of the proposed framework's output over a range of $K$ values and the effects of the merging \& replacing process. 
The $K$ value indicates the maximum fan-ins of the cut circuits; hence, the maximum $K$ value is set to 4, which corresponds to the maximum fan-ins of the cells in our cell library.
Table \ref{tab:diff} shows the average results for 15 benchmark circuits listed in Table \ref{tab:result}, when the framework sets $K$ value to 3 or 4, as well as their corresponding result without merging \& replacing process. 
As discussed in Section\ref{sec:framework}, our developed technology mapper focuses on improving four critical parameters in SFQ circuits including logical depth, \#JJs, PND and \#(DFF+INV). 
In general, \#DFFs metric is used to determine the overhead in the path-depth balancing. Due to the fact that our post-process algorithm substitutes the DFF for INV to achieve path-depth balancing under the permissible conditions. \#(DFF+INV) substitutes \#DFFs as a metric in our evaluation. The framework with $K=4$ improves all the metrics compared with $K=3$. This is the reason that the result set from the cut searching algorithm with a larger $K$ value is the super-set of the lower $K$ value cut circuit sets. Specifically, it is capable of discovering additional cut circuits for the root node and regenerating a representative cut circuit with a lower PND value.
The merging \& replacing process achieves comparable performance under the same logic result condition since DFFs with higher \#JJs are replaced with INVs that require less \#JJs, and the DFFs and INVs in the branches are merged to the root. This operation, on average, reduces \#JJs by 24.88\%. 
% In addition, the Quine–McCluskey algorithm solves the NP-complete problems \cite{umans2006complexity}, which denotes the running time of the Quine–McCluskey algorithm grows exponentially with the number of variables. 

\textbf{2) Comparison with the state-of-the-art technology mapper:} 
Table \ref{tab:result} shows the experimental results for our framework and two baseline mappers. The first baseline mapper is the input netlist $\mathcal{B}_1$ applying our proposed SFQ unique characteristics modifications in Algorithm \ref{alg:pre-process} to insert the DFFs and splitters. The second baseline is SFQmap $\mathcal{B}_2$ \cite{pasandi2018sfqmap}, whose $K$ value is set to 3 to maintain consistency with original works. We unify cell library for all mappers with the cell library presented in Section \ref{sec:cell} to ensure fairness. The same metrics in the $K$ value test are evaluated in this experiment.
Our framework provides considerable improvements on all these critical parameters in all the benchmark circuits compared with $\mathcal{B}_1$. The logical Depth, \#JJs, PND value, and \#(DFFs+INVs) are reduced 32.8\%, 60.5\%, 72.5\% and 53.1\% on average, respectively. It is mainly because $\mathcal{B}_1$ does not provide an efficient mapping algorithm to reduce the overhead. 

In general, our framework $\mathcal{F}$ outperforms baseline SFQmap $\mathcal{B}_2$ in terms of \#JJs, PND value, and the sum of \#DFFs and \#INVs. This verifies our expectation that the proposed framework can reduce the number of Josephson Junctions (\#JJs) and the product of total \#JJs with logical depth (PND). On average, $\mathcal{F}$ reduces the logical depth of all benchmark circuits by 8.3\% as compared to $\mathcal{B}_2$, the DFFs number and logical depth oriented optimization mapping algorithm. 
This demonstrates the advantage of our algorithm, which incorporates MAJ gates and regenerates the cut circuits based on the logic level of leaf nodes in the original circuit rather than simply minimizing the critical paths of the cut circuits with supergates. 
In circuit b11, $\mathcal{F}$ produces a mapping result that has a slightly greater logical depth than $\mathcal{B}_2$. This indicates that $\mathcal{F}$ as a multiple objective oriented optimization mapper, which engages in finding the mapping result by considering overall metrics. 
Therefore, it can expense logical depth in exchange for a result with a lower PND value. While it produces a result with a higher logical depth value, the PND value and \#JJs are reduced significantly in b11.

\begin{table}[h]
    \centering
    \caption{Experimental Results for proposed framework $\mathcal{F}$ ($K=4^-$) and baseline ($\mathcal{B}_2:$ SFQmap \cite{pasandi2018sfqmap})}
    \label{tab:4-SFQmap}

    \begin{tabular}
    {m{1.0cm}<{\centering} m{1.2cm}<{\centering} m{0.9cm}<{\centering} m{1.5cm}<{\centering} m{1.8cm}<{\centering} }
    \hline\hline
    mapper              & Logical Depth    & \#JJs            & PND($\times10^5$) &\#(DFFs+INVs)      \\\hline 
    $\mathcal{F}$             &44.8            &106253            &62.42              &10943              \\[0.8ex]\hline 
    $\mathcal{B}_2:$    &48.8            &123212            &79.10              &13017              \\[0.8ex]\hline
    Imp(Avg)           &$\uparrow$8.2\% &$\uparrow$13.7\%  &$\uparrow$21.1\%   &$\uparrow$15.9\%   \\[0.8ex]\hline
    \end{tabular}
\end{table}

To analyze whether the observed improvements in the comparison with the state-of-the-art technology mapper are primarily contributed by our mapping algorithm, we disable the merging \& replacing in $\mathcal{F}$ and then compare it with SFQmap $\mathcal{B}_2$ again. Due to the merging \& replacing operation can also contribute to the improvement in terms of \#JJs reduction, that is illustrated in the $K$ values test. 
Table \ref{tab:4-SFQmap} summarizes experimental results by average over the preceding 15 benchmark circuits. Although the optimizations in \#JJs are suppressed without the merging \& replacing process, all other critical metrics, such as logical depth, PND value and \#(DFFs+INVs) continue to improve significantly.

% However, this method would lead us to believe that it is responsible for all of the improvements observed in comparisons to baseline mappers.
% Therefore, we disable the merging \& replacing process in $\mathcal{F}$ and then compare it with SFQmap $\mathcal{B}_2$ again.

\section{Conclusion}
In this paper, we proposed a comprehensive logic synthesis framework that is capable of mapping a netlist to an optimal netlist compatible with the SFQ technology, and also proposed MAJ gate circuit configuration in SFQ circuits and applied it into our framework. 
Our proposed mapping optimization algorithm has proved its superiority against the baseline approaches in terms of reducing the overhead associated with path-depth balancing and product of Josephson Junction number with logical depth (PND) for all the tested benchmark circuits, thereby mitigating the non-idealities in these SFQ circuits. Experimental results compared with the state-of-the-art technology mapper indicates that our framework reduces the logical depth, \#(DFFs+INVs), \#JJs, and PND by an average of 8.2\%, 17.5\%, 35.0\%, and 43.1\%, respectively over 15 benchmark circuits.

\clearpage

\bibliographystyle{IEEEtran}
\bibliography{references}

\end{document}